\documentclass[12pt,preprint]{aastex}
\usepackage{graphics,graphicx,rotating}

\newcommand{\dt}{d\theta}

\newcommand{\paren}[1]{\left( #1 \right)}

\newcommand{\bbeta}{\mbox{\boldmath $\beta$}}
\newcommand{\btheta}{\mbox{\boldmath $\theta$}}
\newcommand{\bnabla}{\mbox{\boldmath $\nabla$}}

\newcommand{\rh}{\mbox{$r_h$}}
\newcommand{\rg}{\mbox{$r_g$}}

\shorttitle{}
\shortauthors{Okura, Umetsu, \& Futamase}

\begin{document}

\title{A New Method for measuring Weak Gravitational Lensing Shear using Higher Order Spin-2 HOLICs}
\author{Yuki Okura\altaffilmark{1}} 
\email{aepgstrx@astr.tohoku.ac.jp}

\author{Toshifumi Futamase\altaffilmark{1}}
\email{tof@astr.tohoku.ac.jp}

\altaffiltext{1}
 {Astronomical Institute, Tohoku University, Sendai 980-8578, Japan}

\begin{abstract}
We investigate the possibility to use higher order moments of gravitational lensed images in the weak lensing analysis.
For this purpose we employ spin-2 component of HOLICs(Higher Order Lensing Characteristics) developed by us.
We test the weak lensing analysis with spin-2 HOLICs using actual, ground based Subaru observations of the 
massive galaxy cluster A1689(z=0.183). It turns out that spin-2 HOLICs of order up to 8 are sufficiently applicable 
for weak lensing analysis after correcting 
PSF anisotropy as well as isotropic PSF smearing.

\end{abstract}

\section{Introduction}

Weak gravitational lensing is now regarded as a powerful tool to reveal the distribution of dark matter 
of clusters of galaxies but also large scale structure. 
It also useful to measure the equation of state parameter of dark energy by observing cosmic shear.  

In the usual treatment of the weak lensing analysis, the quadrupole moment of background galaxy images is 
used to quantify the image ellipticity.
In recent years, there have been theoretical efforts to include the next order distortion effects as well as 
the usual quadrupole distortion effect in the weak lensing analysis (Goldberg \& Natarajan 2002; 
Goldberg \& Bacon 2005; Bacon et al. 2006; Irwin \& Shmakova 2006; Goldberg \& Leonard 2007).
We have proposed to use some combinations of octopole/higher multipole moments of background images with definite spin properties 
which we call the Higher Order Lensing Image's Characteristics (HOLICs), 
and have shown that there is simple relations between HOLICs with spin-1 (spin-3) 
and the third-order weak lensing effects, or gravitational flexion with spin-1 (spin-3)(Goldberg \& Bacon 2005), 
and thus HOLICs serve as a direct measure for the flexion(Okura, Umetsu \& Futamase 2007, 
In the following we refer this paper as OUF2007).
We have also developed realistic method for measuring flexion by the HOLICs approach by fully extended the KSB formalism to take into account higher-order PSF anisotropy as well as isotropic PSF smearing (Okura, Umetsu \& Futamase 2008. 
In the following we refer this paper as OUF2008).
The method has been applied to actual, ground-based Subaru observation of A 1689 and has able to obtain a bimodal feature in the central region of the cluster which is not seen in the usual quadrupole weak lensing analysis.

In this paper we consider a possibility to apply HOLICs method to measure shear, particularly cosmic shear. 
Weak lensing shear analysis has suffered from various sort of noise. Here we concern the intrinsic noise which 
is nothing to do with observational condition and method of shape measurement.  
The intrinsic noise can be reduced by averaging a sufficient number of background images.  
Since cosmic shear induces at most of the order of 1\% distortion, it is absolutely necessary to reduce any source of noises. 
Thus  not only large number density of background images but also wide area of survey is necessary for the cosmic shear analysis 
which is not always possible in the actual observation. 
Therefore it would be very useful to have a method to reduce the intrinsic noise. 
We will try to do this by using information of high order, non-dimension and spin-2 shapes which are called as "spin-2 HOLICs".
The shear distortion has no dimension and is spin-2, thus the shear affects strongly not only complex ellipticity 
but also spin-2 HOLICs.
From the differences of inner and outer region of image, spin-2 HOLICs of different order have different intrinsic noises 
at least partially. In this way the intrinsic noise can be reduced by averaging spin-2 HOLICs of different orders.
In this paper, we define spin-2 HOLICs of various order, and calculate the reduced shear and PSF correction. 
Then we evaluate the relative precision of spin-2 HOLICs of different order by testing them using STEP 1 
and find that Spin-2 HOLICs up to order 8 (which is a combination of $2^8$ multipole moments of shape) are useful 
in the shear analysis. Finally we apply the method to Abell 1689 cluster.

\section{Basis of Weak Lensing}
The gravitational deflection of light rays can be described by the
lens equation,
\begin{equation}
\label{eq:lenseq}
\bbeta = \btheta - \bnabla \psi(\btheta),
\end{equation}
where $\btheta$ and $\bbeta$ are the angular positions of the image and source, respectively, 
and $\psi(\btheta)$ is the effective lensing potential, which
is defined by the two-dimensional Poisson equation as
$\nabla^2\psi(\btheta)=2 \kappa(\btheta)$, with the lensing convergence.
Here the convergence $\kappa=\Sigma_m \Sigma_{\rm crit}^{-1}$ is the
dimensionless surface mass density projected on the sky, normalized with
respect to the critical surface mass density of gravitational lensing
$\Sigma_{\rm crit} = (c^2D_s)/(4\pi GD_d D_{ds})$,
where $D_d$, $D_s$, and $D_{ds}$ are the angular diameter distances
from the observer to the deflector, from the observer to the source,
and from the deflector to the source, respectively.
By introducing the complex gradient operator, $\partial = \partial_1 +
i\partial_2$ that transforms as a vector, 
$\partial'=\partial e^{i\phi}$, with $\phi$ being the angle of rotation,  
the lensing convergence $\kappa$ is expressed as
\begin{equation}
\label{eq:kappa}
\kappa = \frac{1}{2}\partial\partial^* \psi,
\end{equation}
where $^*$ denotes the complex conjugate.
Similarly, the complex gravitational shear of spin-2 is defined as
\begin{equation}
\gamma \equiv \gamma_1+i\gamma_2= \frac{1}{2}\partial\partial \psi.
\end{equation} 
Note that a quantity is said to have spin-s if it has the same value 
after rotation by $2\pi/s$.  

\section{HOLICs}
HOLICs are introduced by us(OUF2007) as particular 
combinations with a definite spin of multipole moments 
of images. Here we review briefly the HOLICs formalism.
It is very useful to use complex moment instead of 
usual moment of images to define HOLICs(OUF2008).  
 
First, we define complex displacement as
\begin{eqnarray}
X^1_1&\equiv&d\theta_1+id\theta_2\\
X^N_M &\equiv&(\dt_1 + i\dt_2)^{(N+M)/2}(\dt_1 - i\dt_2)^{(N-M)/2},
\end{eqnarray}
where $d\theta_i$ means displacement vector from image center and $N$ and $M$ means order of $d\theta$ and spin number, respectively.
Similarly, we define $Y^N_M$ as complex displacement of source image 
defined by the displacement vector $(d\beta_i)$ from the source center.

Complex moments of images having brightness distribution $I(\btheta)$
are defined as follows.
\begin{equation}
Z^N_M\equiv \int d^2\theta X^N_M q[I(\btheta)],
\end{equation} 
where $q[I(\btheta)]$ is an appropriate weight function and is taken as the brightness distribution $I(\btheta)$ itself in this section for simplicity.  
In this notation, the spin-2 HOLICs of order N is defined as $Z^N_2/Z^N_0$ (N is an even number).
For example, N=2 is complex ellipticity $\chi$, and N=4 is $\eta$ and N=6 is $\upsilon_{II}$ 
according to the notation in OUF2008. 
Similarly, we define $Z^{N(s)}_M$ as complex moments of the source image using $d\beta_i$ and $Y^N_M$.

In the shear dominant field, the local expanded the lens equation is expressed using the complex moments as
\begin{equation}
Y^1_1=(1-\kappa)\left(X^1_1-gX^{1*}_1\right),
\end{equation}
where asterisk $^*$ means complex conjugate and $g$ is reduced shear $\gamma/(1-\kappa)$.
Using the above form of the lens equation we have the following simple relation between the intrinsic and the 
observed spin-2 HOLICs. 
\begin{equation}
\frac{Z^{(s)N}_2}{Z^{(s)N}_0}\approx\frac{Z^N_2}{Z^N_0}-\frac{N+2}{2}g.
\end{equation}

If we assume the average of intrinsic spin-2 HOLICs to vanish, 
the reduced shear is obtained directly by observing spin-2 HOLICs as
\begin{equation}
g\approx\left<\frac{2}{N+2}\frac{Z^N_2}{Z^N_0}\right>.
\end{equation}
Since the maximum value of absolute spin-2 HOLICs are 1,
the maximum value of $|g|$ is $2/(2+N)$ in the this approximation .

\section{Weighted HOLICs and PSF correction for real observation}
In the actual observation, the data have random noise and images are smeared by isotropic PSF and distorted by anisotropic PSF. 
In order to correct these effects, we introduce weighted HOLICs and PSF correction. 
The general treatment of the PSF correction may be found, for example, 
in Kaiser, Squires, \& Broadhurst 1995 and Bartellman \& Schneider 2001.
The technical details of the PSF corrections in measuring HOLICs is found in OUF2008. 

We redefine the weighted HOLICs by using an extra weight function $W(X^2_0/\sigma^2)$ with 
a characteristic scale $\sigma$ as
\begin{eqnarray}
Z^N_M\equiv\int d^2\theta I(\btheta)X^N_M W\left(\frac{X^2_0}{\sigma^2}\right) 
\end{eqnarray}
The weighted HOLICs of source $Z^{(s)N}_M$ are defined similarly in the source plane using $d\beta_i$ and $Y^N_M$. 
Using the lens equation as well as the expansion of the weight function 
\begin{eqnarray}
W\left(\frac{Y^2_0}{\sigma^{s\, 2}}\right)=W\left(\frac{X^2_0-Re[\delta^* X^2_2]}{\sigma^2}\right)
\approx W\left(\frac{X^2_0}{\sigma^2}\right) -\frac{Re[\delta^* X^2_2]}{\sigma^2} W'\left(\frac{X^2_0}{\sigma^2}\right),
\end{eqnarray}
the reduced shear effects for spin-2 weighted HOLICs becomes 
\begin{eqnarray}
\frac{Z^{(s)N}_2}{Z^{(s)N}_0}\approx\frac{Z^N_2}{Z^N_0}-\paren{\frac{N+2}{2}+\frac{{Z^{N+2}_0}'}{\sigma^2Z^N_0}}g-\paren{\frac{N-2}{2}\frac{Z^N_4}{Z^N_0}+\frac{{Z^{N+2}_4}'}{\sigma^2Z^N_0}}g^*\equiv \frac{Z^N_2}{Z^N_0}-C^N_0g-C^N_4g^*,
\end{eqnarray}
where HOLICs with $'$ are measured with $W'(x)=\partial W(x)/\partial x$ instead of $W(x)$.

Next, the anisotropic PSF is corrected as
\begin{eqnarray}
\frac{Z^{Niso}_2}{Z^{Niso}_0}\approx\frac{Z^{Nobs}_2}{Z^{Nobs}_0}-P^N_0\chi_q-P^N_4\chi^*_q,
\end{eqnarray}
where
\begin{eqnarray}
P^N_0&\equiv& \frac 1{Z^N_0} {\left( \frac{    N(N+2)}{8}Z^{N-2}_0+\frac{(N+2)}{2\sigma^2}{Z^{N}_0}'+\frac{1}{2\sigma^4}{Z^{N+2}_0}'' \right)}\\
P^N_4&\equiv& \frac 1{Z^N_0} {\left( \frac{(N-2)(N-4)}{8}Z^{N-2}_4+\frac{(N-2)}{2\sigma^2}{Z^{N}_4}'+\frac{1}{2\sigma^4}{Z^{N+2}_4}'' \right)},
\end{eqnarray}
where the quantities with upper subscript "obs" and "iso" mean the quantities before and after making the anisotropic PSF correction, respectively, and HOLICs with $''$ are measured with $W''(x)=\partial^2 W(x)/\partial^2 x$ instead of $W(x)$.  
In the  actual observation, higher order corrections using higher order spin quantities such as $P^N_6$(spin-6) might be necessary for accurate shape measurement. 
However one can see in next section that the constructed mass distribution is accurate enough 
even in the approximation using up to $P^N_4$. 

The $\chi_q$ is spin-2 anisotropic component of PSF (OUF2008), 
and obtained from HOLICs measured from star images as follows, 
\begin{eqnarray}
\chi_q\approx\frac{1}{P^N_0}\frac{Z^{Nobs}_{2star}}{Z^{Nobs}_{0star}}.
\end{eqnarray}
Finally, we obtain the relation between spin-2 HOLICs and reduced shear from each order of HOLICs after isotropic PSF correction as follows, 
\begin{eqnarray}
\frac{Z^{Niso}_2}{Z^{Niso}_0}\approx\paren{C^N_0-\paren{\frac{C^N_{0 star}}{P^N_{0 star}}}P^N_0}g-\paren{C^N_4-\paren{\frac{C^N_{0 star}}{P^N_{0 star}}}P^N_4}g^*
%g&\approx&\left< \frac{1}{C^N_2-\paren{\frac{C^N_{2 star}}{P^N_{2 star}}}P^N_2}\frac{Z^{Niso}_2}{Z^{Niso}_0} \right>\\
%C^N_2&\equiv&\paren{\frac{N+2}{2}+\frac{1}{\sigma^2}\frac{{Z^{N+2}_0}'}{Z^N_0}}
\end{eqnarray}

\begin{figure*} 
\epsscale{1.0}
\plotone{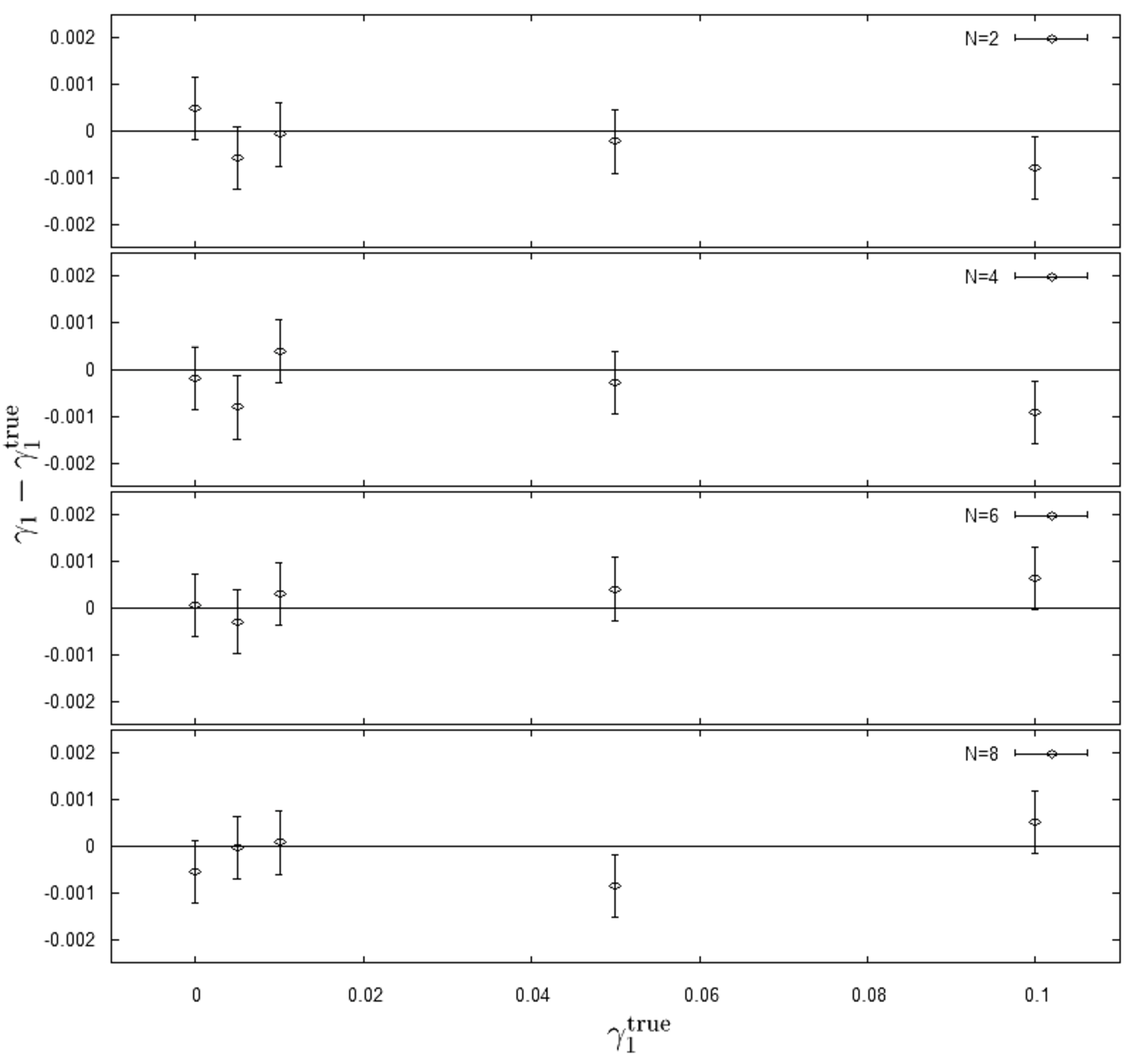}
\caption{
\label{fig:STEP1}
The results of analysis of STEP1 simulations in each order of HOLICs shear measurement.
The horizontal axis are the true(input) shear, and the vertical axis are the difference of estimated and true shear with 1$\sigma$ errors.
Errors of estimated shear are same levels and less than 1$\sigma$ error in All order.
} 
\end{figure*}

We have tested our shear estimation by HOLICs using the ready-made simulation from STEP, 
and have obtained the input value of the shear when we use higher order PSF corrections
($C^N_0$ and $C^N_4$) 
in all order of HOLICs measurement and do not use faint images in the higher order HOLICs shear measurement. 
This can be seen in Figure \ref{fig:STEP1} where the results of analysis of STEP1 simulations are shown in each 
order of HOLICs shear measurement.
In this simulation we used about $10^5$ brighter images with $20.3<$MAG$<23.3$ and $8.6$ arcmin $^{-2}$.
In order to use fainter images for higher order HOLICs shear measurement we will need higher order polarization matrices 
to correct PSF appropriate for higher order HOLICs shear measurement which will be presented in the forthcoming paper.

\section{A1689 analysis}
In this section, we show the demonstration of applying the weak lensing analysis using spin-2 HOLICs with order $N=2,4,6,8$ to 
Subaru imaging observations of the massive galaxy cluster Abell 1689 at $z=0.183$. Abell 1689 is one of 
the best studied clusters(e.g., Tyson \& Fisher 1995; King, Clowe, \& Schneider et al; Bardeau et al.; 
Broadhurst et al. 2005a; Broadhurst et al 2005b; Halkola et al. 2006; Leonard et al. 2007; Limousin et al. 2007; 
Umetsu, Broadhurst, Takada 2007; Umetsu \& Broadhurst 2007), and therefore serves as an ideal target for 
testing new method.   

We use Subaru Suprime-Cam $i'$-band data
which has $30' \times 25'$ field and $0".202$pixel$^{-1}$.
We analyzed 3000 $\times$ 3000 pixel(or about $10' \times 10'$)which is center of the field,
and the seeing is $0".88$ by FWHM.

We used IMCAT(Kaiser et al. 1995) and some scripts (K. Umetsu, private communication) for image detection 
{\bf and measure position, gaussian radius "$\rg$", half-light radius "$\rh$" and magunitude of detected objects in some stage of analysis pipeline.}

\begin{figure*} 
\epsscale{1.0}
\plotone{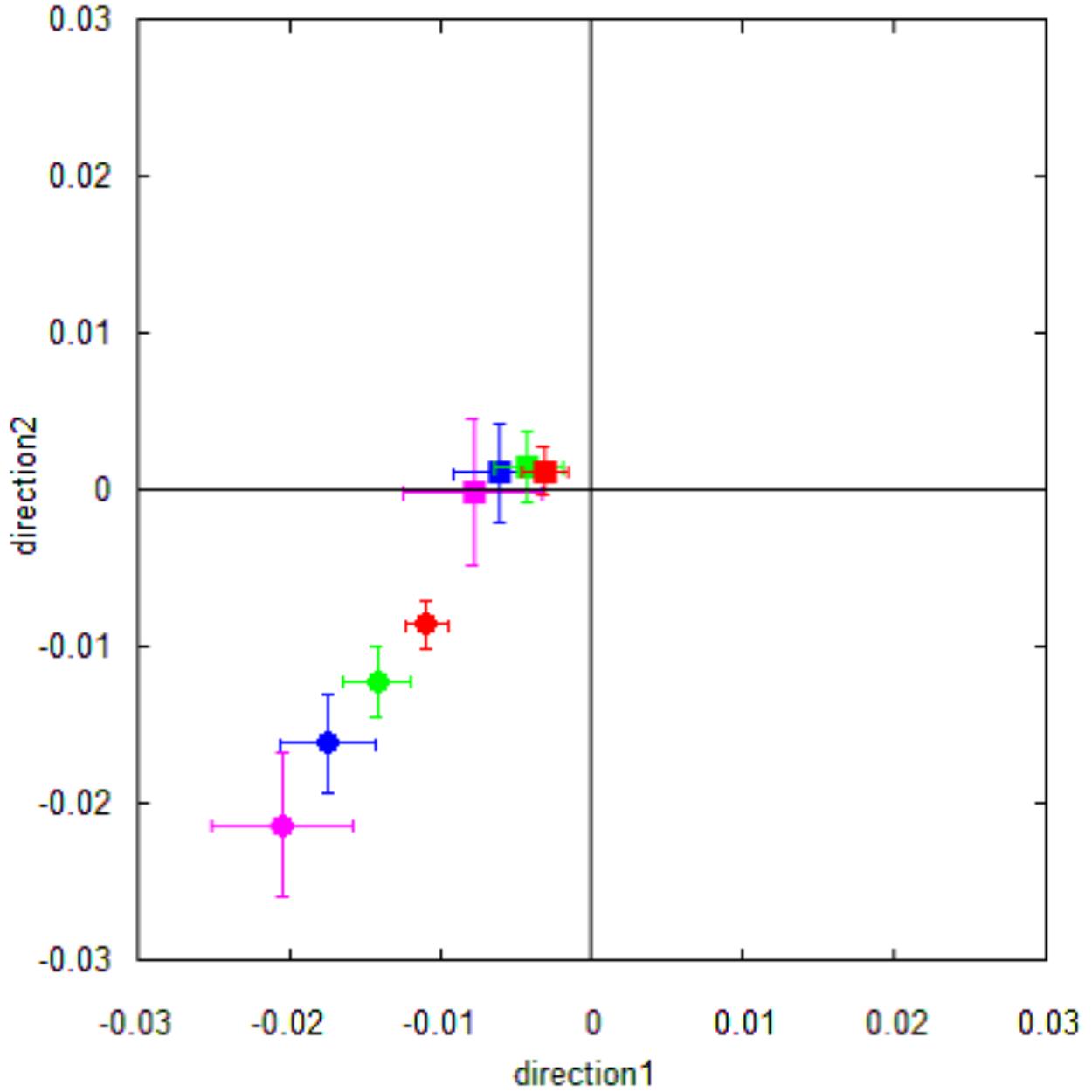}
\caption{
\label{fig:PSF}
The plots are the average of observed and aniso-PSF corrected Spin-2 HOLICs.
Red, green, blue and purple are N=2,4,6 and 8 respectively,
and circles are observed squares is aniso-PSF corrected Spin-2 HOLICs.
We can see anisotropic PSF of all order of Spin-2 HOLICs were corrected.
} 
\end{figure*}

We used 81 star images which have $2.2<\rh<2.5$ and $21<$MAG$<22.5$ for PSF correction,
and we used 3366 galaxies(about 33 arcmin$^{-2}$) with $2.5<\rh<10$ and $22<$MAG$<25.5$.
Figure \ref{fig:PSF} is the plots of the average of spin-2 HOLICs before and after anisotropic PSF correction.

\begin{figure*} 
\epsscale{1.0}
\plotone{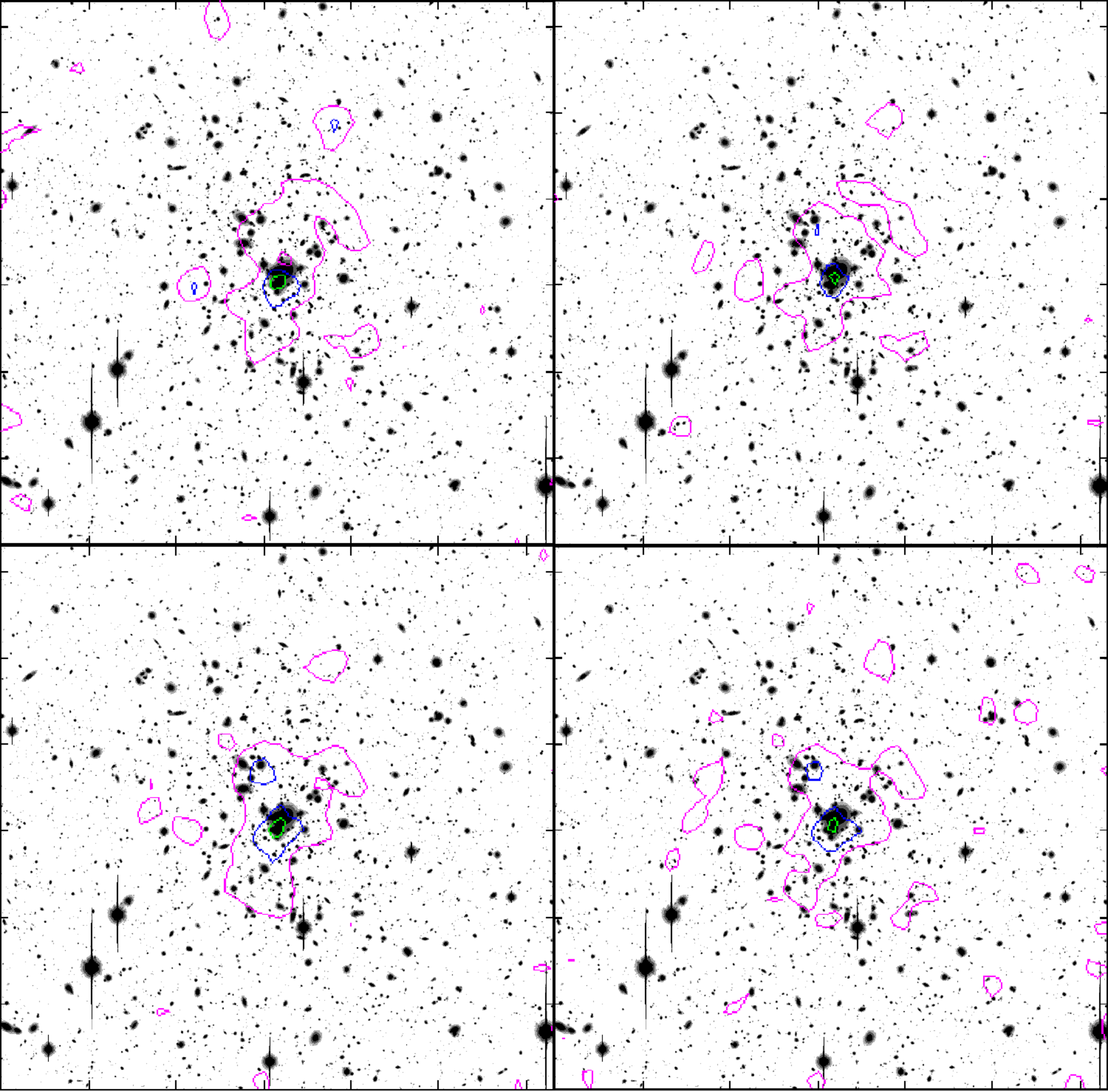}
\caption{
\label{fig:A1689}
The reconstructed mass distribution in the central $10'\times 10'$ region of A1689 
by spin-2 HOLICs of various order up to 8. 
{\bf The contours are spaced in units of $0.2 (\approx 3\sigma)$,
therefore purple ,blue and green lines mean $\kappa=0.2, 0.4$ and $0.6$ respectively.}
North is to the top, and East to the left.
} 
\end{figure*}

\begin{figure*} 
\epsscale{1.0}
\plotone{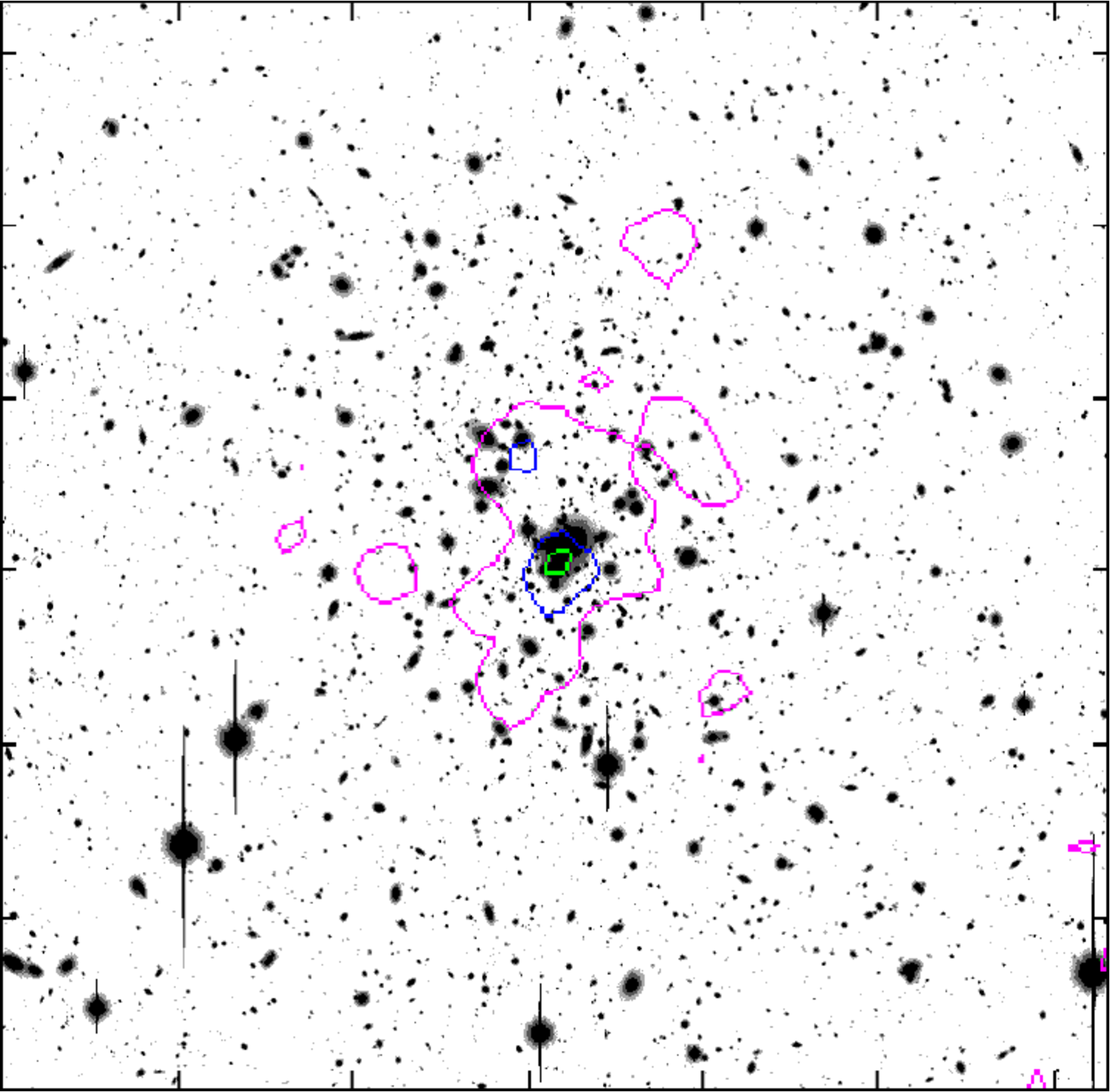}
\caption{
\label{fig:A1689mix}
The averaged mass distribution in the same region of Fig.\ref{fig:A1689} over that measured by spin-2 HOLICs of order N up to 8.  
} 
\end{figure*}

\begin{figure*} 
\epsscale{1.0}
\plotone{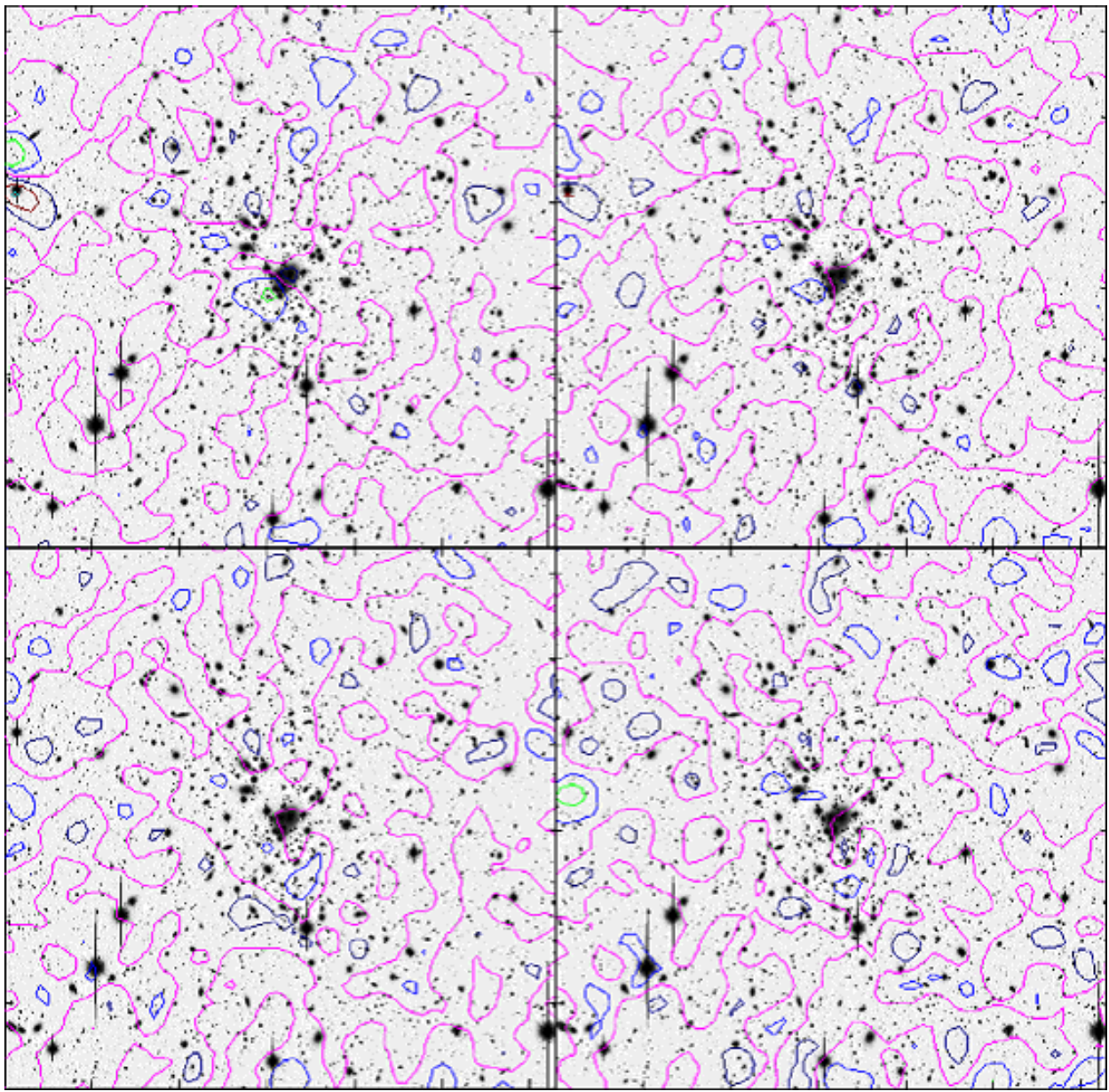}
\caption{
\label{fig:A1689B}
The B-mode in the central $10'\times 10'$ region of A1689 
by spin-2 HOLICs of various order up to 8. 
The contours are spaced in units of $1\sigma (\approx 0.2)$,
and purple is 0, blue is 0.2 and so on.
} 
\end{figure*}

\begin{figure*} 
\epsscale{1.0}
\plotone{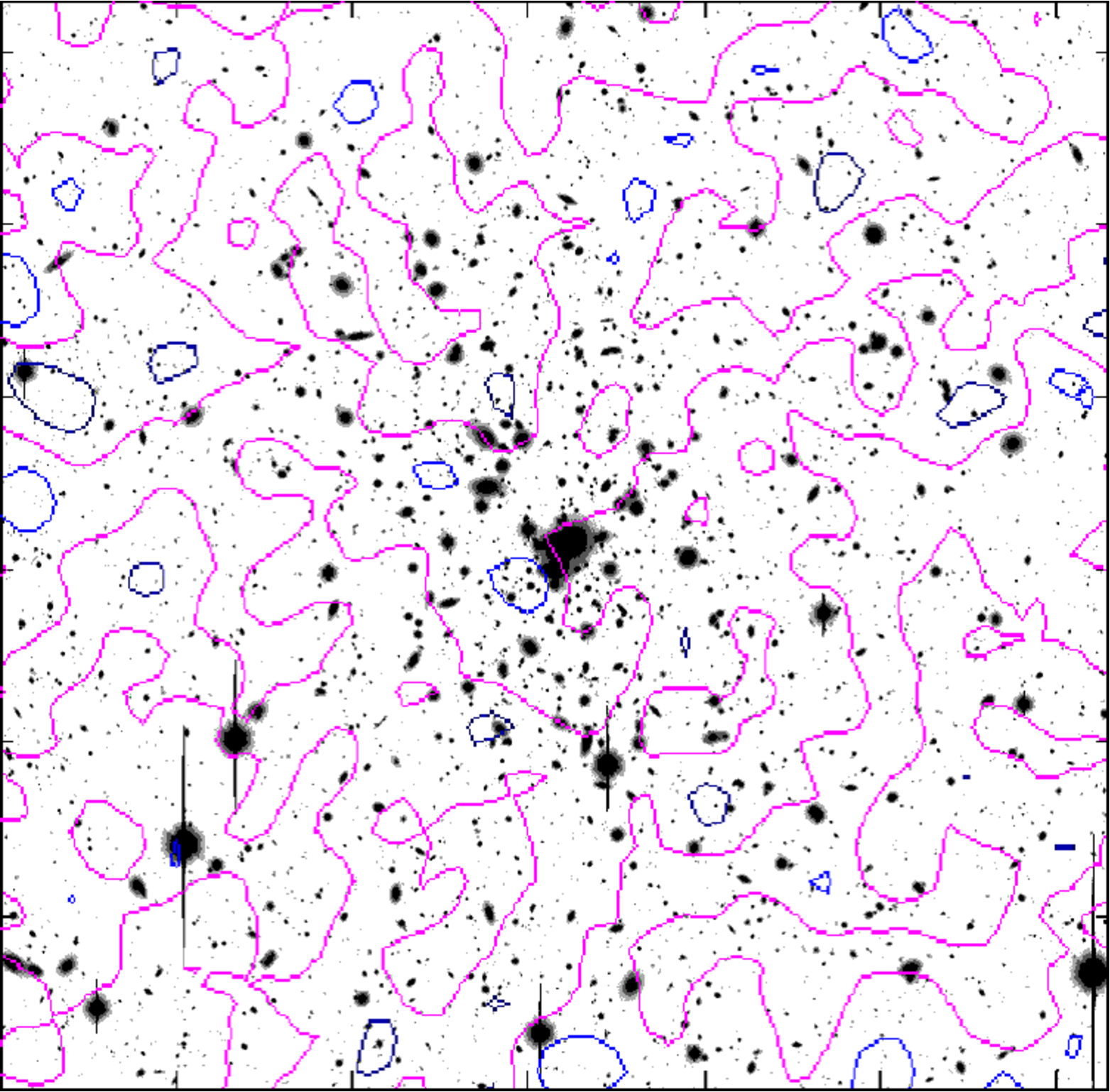}
\caption{
\label{fig:A1689Bmix}
The averaged mass distribution in the same region of Fig.\ref{fig:A1689B} over that measured by spin-2 HOLICs of order N up to 8.  
The contours are spaced in units of $1\sigma (\approx 0.2)$,
and purple is 0, blue is 0.2 and so on.
} 
\end{figure*}

Figure \ref{fig:A1689} are the reconstructed mass distributions measured with HOLICs of order $N=2, 4, 6, 8$. 
The smoothing is $0".15$ by gaussian radius and the interval of successive contours is $\Delta \kappa=0.2$ 
and the lowest contour is $\kappa=0.2$.
Reconstructed B-modes of each order have dispersion $1\sigma\approx0.065, 0.060, 0.062, 0.073$, respectively.
Figure \ref{fig:A1689mix} is the average of these four reconstructed mass distribution and 
the interval of successive contours is $\Delta \kappa=0.2$ and the lowest contour is $\kappa=0.2$.
Figure \ref{fig:A1689B} are the B-mode of Abell 1689 measured with HOLICs of order $N=2, 4, 6, 8$,
the mapping status is same as figure \ref{fig:A1689}.
The contours separate $\kappa_B=0.2$ and purple is $\kappa_B=0$.
These figures show that the absolute values of B-mode are less than 0.2 in almost fields.
Figure \ref{fig:A1689Bmix} is the average of four B-mode,
the contours separate $\kappa_B=0.2$ and purple is $\kappa_B=0$.
Averaged B-modes has dispersion $1\sigma\approx0.059$.

These results clearly show that we can reconstruct A1689 mass distribution using the higher order spin-2 HOLICs 
as well as complex ellipticity $\chi$. The dispersion of B-mode is reduced by combining B-mode of each order.
Compared with the case of using only complex ellipticity $\chi$, the dispersion is reduced by about 10\%.
{\bf We can even detect a stronger sub clump ($\kappa>0.4$) by using higher order spin-2 HOLICs than by using 
only complex ellipticity.}
%We can even detect a sub clump by using higher order spin-2 HOLICs which is not detected by using 
%only complex ellipticity. 
We can see the suppression of the noise in the averaged mass distribution.  

The detail analysis of Abell 1689 (e.g. radial profiles and so on) will be presented in the forthcoming paper.

\section{Conclusion}
We examined the possibility to use higher order multipole moments to measure the weak shear. 
For this purpose we use spin-2 HOLICs of order N larger than 2. 
In order to apply the actual observational data, we introduced 
the weighted HOLICs and investigated PSF correction. 
Then we apply the method to the massive galaxy cluster A1689(z=0.183), and  found surprisingly 
that spin-2 HOLICs of order N up to 8 may be able to measure the mass distribution with 
sufficient accuracy. 
The reconstructed mass distributions from higher order spin-2 HOLICs show a 2nd peak in the central region, 
and the dispersions of B-mode measured by these HOLICs are of the order of that measured by the usual complex ellipticity. 
The combined B-mode is reduced by about 10\% and thus the accuracy of the combined mass distribution is improved by about 10\% over that measured only by the complex ellipticity. 

\begin{figure*} 
\epsscale{1.0}
\plotone{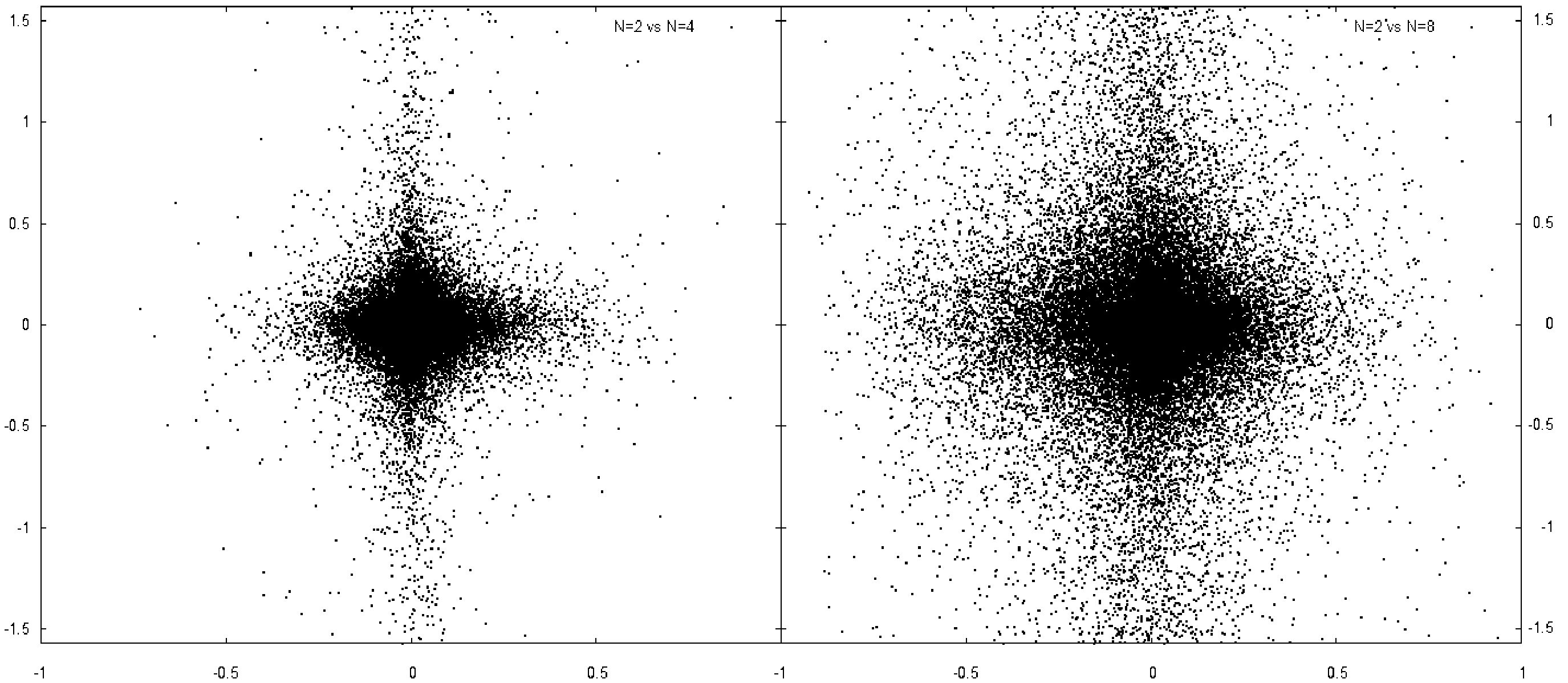}
\caption{
Figures show the differences between spin-2 HOLICs of order $N=4$(left), 8(right) and complex ellipticity $\chi$.
The horizontal axis is the differences between the absolute value of spin-2 HOLICs of order N=4(8) and absolute value of complex ellipticity.
The vertical axis is the angles determined by spin-2 HOLICs of order N=4(8) relative to the direction determined by complex ellipticity.
} 
\label{fig:dH2}
\end{figure*}

The improvement is obtained by the partial independence of the shear measured by spin-2 HOLICs of different orders. The higher the order, the outer the region of the image the shear is measured.  
Thus spin-2 HOLICs of different order measure the shear in the different region of 
the image. 
In general the direction determined by the measured shear near the central region of the image 
is not necessarily the same with the direction determined by the measured shear in the outer part. 
This is the basic reason why we can improve the accuracy of the measurement of mass distribution. 
Of course the weight function of the different order of multipoles are overlapped each other, and thus 
the shear measured by them are only partly independent. 
We show in Figure \ref{fig:dH2} the difference between 
the directions of shear determined by spin-2 HOLICs of different orders as  a function of the difference between the magnitudes of the shear determined 
by spin-2 HOLICs of different orders.
We show this figure because  the difference is important to reduce intrinsic noise, and we will 
address this point more detail in the forthcoming paper.
This Figures are made by analyzing the data obtained in the Subaru imaging observation 
of the blank region ELAIS N1.
We use 35521 galaxies in 0.28deg$^2 \approx$ 2$\times$(12.6)$^2\pi$ arcmin$^2$, corresponding to a mean surface number density of 35.6 arcmin$^{-2}$.
They clearly suggest that the shear measured by spin-2 HOLICs of larger order is more independent 
of that measured by the usual complex ellipticity.

In this letter we have shown the usefulness of the higher order spin-2 HOLICs in the shear measurement.E
This will open a new possibility to improve the accuracy of the weak lensing analysis without doing further observations. 
Although we have tested our method to weak lensing analysis for cluster of galaxies, 
we can also apply the method to the cosmic shear. It is very interesting to see how we can improve the measurement of the cosmological parameters by the cosmic shear using higher order spin-2 HOLICs. 

For this purpose, we need more detailed analysis of this method, for example the test of other PSF data of STEP I and STEP II.
These will be given in the forthcoming paper.

\acknowledgments
We thank K. Umetsu for useful discussion and providing his scripts for the image detection 
, T. Yamada for the observation of N1 and T. Okamura for useful discussion. 
The work is partially supported by Research Fellowships of the Japan Society for the Promotion of
Science for Young Scientists for YO. This work is also supported in part by a Grants-in-Aid for Scientific Research from JSPS (Nos. 18072001, 20540245 for TF) as well as by Core-to-Core Program "International Research Network for Dark Energy".

\newpage



\begin{thebibliography}{99}
\bibitem{Bacon06}
	Bacon, D. J., Goldberg, D. M., Rowe, B. T. P., \& Taylor, A. N. 2006, \mnras, 365, 414

\bibitem{BARD1689}
	Bardeau, S. et al. 2005 A\&A, 434, 433 
	
\bibitem{review} 
	Bartelmann, M., \& Schneider, P. 2001, Phys.Rep., 340, 291

\bibitem{B05b} 
	Broadhurst, T. et al. 2005, \apj, 621, 53  (Broadhurst et al. 2005b)

\bibitem{B05a}
	Broadhurst, T., 
	Takada, M., 
	Umetsu, K.,
	Kong, X., 
	Arimoto, N., 
	Chiba, M.,  \& Futamase, T. 2005, 619,
	143L (Broadhurst et al. 2005a)
	
\bibitem{GN02}
	Goldberg, D.~M. \& Natarajan, P. 2002, \apj, 564, 65
	
\bibitem{Flexion} 
	Goldberg, D. M., \& Bacon, D. J. 2005, ApJ, 619, 741

\bibitem{Flexion2} 
	Goldberg, D. M., \&  Leonard, A. 2006, ApJ, 660, 1003
	
\bibitem{halkola07}
	Halkola, A.
	Seitz, S.,
	\& Pannella, M. 2006, \mnras, 372, 1425
	
\bibitem{Sextupole}
	Irwin, J., \& Shmakova, M. 2006, ApJ, 645, 17
	
\bibitem{IMCAT} Kaiser, N.. Squires, G., Broadhurst, T. 1995, ApJ, 449, 460

\bibitem{KA1689} 
	King, L. J., Clowe, D. I., Schneider, P. 2002 A\&A, 383, 118 
	
\bibitem{GA1689} 
	Leonard, A., Goldberg, D.~M., Haaga, J.~L., 
	Massey, R. 2007, ApJ, 666, 51L
	
\bibitem{SW-A1689}
	Limousin, M. et al. 2007, ApJ, 668, 643
	
\bibitem{HOLICs} 
	Okura, Y., Umetsu, K., \& Futamase, T., 2007, ApJ, 660, 995 
	
\bibitem{HOLICs2} 
	Okura, Y., Umetsu, K., \& Futamase, T., 2008, ApJ, 680, 1.
	
\bibitem{TA1689}
	Tyson, J. A., \& Fisher, P., 1995 ApJ 446 L55
 
\bibitem{UA1689} 
	Umetsu, K., Takada, M., Broadhurst, T. 2007,
	Mod.~Phys.~Lett.~A, 22, 2099 (arXiv:astro-ph/0702096)

\bibitem{Umetsu07}
	Umetsu, K. \& Broadhurst, T. 2007, submitted to ApJ 
	(arXiv:astro-ph/0712.3441)
\end{thebibliography}
\end{document}